\documentclass[aps,prl,twocolumn,showpacs,amsmath,amssymb,groupedaddress,showpacs]{revtex4}
\usepackage{amsfonts}
\usepackage{amsmath}
\usepackage{dcolumn}
\usepackage[dvips]{graphicx,color}
\begin{document}

\title{P-wave Pairing in Two-Component Fermi System with Unequal Population near Feshbach Resonance}
\author{Renyuan Liao}
\author{Florentin Popescu}
\author{Khandker Quader}
\affiliation{Department of Physics, Kent State University, Kent, OH 44242}

\date{\today}

\begin{abstract}

We explore $p$-wave pairing in a single-channel two-component Fermi system with unequal
population near Feshbach resonance. Our analytical and numerical study reveal
a rich superfluid (SF) ground state structure as a function of imbalance.
In addition to the state $\Delta_{\pm 1} \propto Y_{1\pm 1}$, a multitude of ``mixed'' SF states
formed of linear combinations of $Y_{1m}$'s give global energy minimum under a
phase stability condition; these states exhibit variation in energy with
the relative phase between the constituent gap amplitudes. States with local energy
minimum are also obtained.
We provide a geometric representation of the states. A $T$=0 polarization vs. p-wave coupling
phase diagram is constructed across the BEC-BCS regimes. With increased polarization,
the global minimum SF state may undergo a quantum phase transition to the local minimum SF state.

\end{abstract}


\maketitle

{\it Introduction:} Discovery of s-wave paired fermion condensates in cold atoms
\cite{BCSBEC} subjected to s-wave Feshbach resonance (FR) have led to
explorations in several fascinating directions. One is the study of
superfluidity in asymmetrical Fermi systems, a subject of intense
experimental and theoretical research \cite{IMPEX,IMPTH,LQ06}. Another is
the prospect for attaining fermion condensates with pairs having
non-zero relative orbital angular momentum. Recent
observations of p-wave FR \cite{REG03,TIC04,ZHA04,SCH05} in $^6Li$
and $^{40}K$ have raised the possibility of observing p-wave
superfluidity in cold atoms, leading to renewed theoretical
interest. P-wave superfluidity in the BEC-BCS crossover region
has been studied using fermion-boson models applied to  single-component
Fermi gas (where FR occurs in the same hyperfine state)
\cite{GUR05,CHE05}, and using fermion-only models at $T$=0 both in
two-component Fermi gas (where FR occurs between different hyperfine
states) \cite{TIN05} and in single-component case \cite{BOT05}.
Ohashi \cite{OHA05} considered both cases at finite-T. While there
have been fervent theoretical efforts \cite{IMPTH,LQ06} on pairing in asymmetrical
 Fermi systems subject to s-wave FR, work on the corresponding
p-wave case is lacking.

In this paper, within a fermion-only model,
we study p-wave superfluidity near a p-wave FR in a two-component
Fermi system with {\it unequal} populations in two hyperfine states.
It is conceivable that in future, population imbalanced systems under p-wave FR
would be realized similar to the ones subject to s-wave FR. We hope that this work will
motivate experimental efforts.

It has been pointed out \cite{TIN05,TIC04} that unlike liquid $^3He$,
pairing interaction in cold atoms is highly anisotropic in
``spin''-space. (``spin'' referring to hyperfine states). For
example, in $^6Li$, when the hyperfine pair $|1/2,-1/2>$ is at
resonance, the pairs $|1/2,1/2>$ and $|-1/2,-1/2>$ would not be. So,
in studying pairing in this two-component system, intra-spin
interactions need not be considered. Pairs in p-wave superfluids
with unlike ``spin'' components can however have different orbital
angular momentum content, so the gap parameters $\Delta_{lm} \propto$
$\ell=1$ spherical harmonics $Y_{1m}$ ($m= \pm 1,0$).

Our analytic and numerical considerations of the population imbalanced system lead
to several interesting results and predictions:
(i) We find a rich superfluid ground state (GS) structure:
The p-wave state $\Delta_1$(hence $\Delta_{-1}$ by symmetry) give a GS {\it global} minimum energy. Interestingly,
under a specific condition involving the relative phases of the
three pairing amplitudes $\Delta_m$'s, a multitude of ``mixed'' states of the form
$a\Delta_0 +b\Delta_1 +c\Delta_{-1}$ ($\Delta_{1m} \equiv \Delta_m$ here) are degenerate with $\Delta_{\pm 1}$.
In addition, we find states with {\it local} energy minimum.
(ii) We provide a geometric representation of the p-wave superfluid
states (Fig 1):
The states exhibiting global minimum
lie on a ``semicircle''  formed by the intersection of
the surface of the sphere formed by $\Delta_1, \Delta_{-1}, \Delta_0$ with a plane
defined by $|\Delta_1| +|\Delta_{-1}|$ = const.
(iii) We obtain polarization (P) vs p-wave coupling phase diagram at $T$=0. (Fig 2).
The superfluid phase SF1 comprises of states with global minimum, while SF2 of states with the local minimum.
We also find a region of phase separation PS
between SF2 and the normal phase N.
In this two-component system, PS persists onto full polarization.
(iv) These raise the intriguing possibility for a $T$=0 quantum phase transition
from SF1 to SF2 at finite polarization (Fig 2).
Additionally, transitions at finite-T would also occur.
(v) In the limit $P \rightarrow$ 0, we find that the ground state
structure of $P \ne$ 0, is preserved, hence
richer than that  obtained earlier \cite{TIN05}.

{\it Model:} We consider a two-component Fermi system with unequal ``spin''
($\uparrow, \downarrow$) population, but equal masses for
the unlike fermions. We take the interaction between unlike fermions
to be isotropic in orbital space. Following the rationale above,
interactions between like fermions are taken to be zero. Since we
adjust self-consistently the chemical potential with the strength
and sign of the coupling, in our fermion-only model, molecules would
naturally appear as 2-fermion bound states. The pairing Hamiltonian
is then given by:
\begin{eqnarray}\label{Ham}
{\mathcal{H}}&=&\sum_{\mathbf{k}\sigma}\xi_{\mathbf{k}\sigma}c^{\dag}_{\mathbf{k}\sigma}c_{\mathbf{k}\sigma}\nonumber\\
&+&\sum_{\mathbf{k}\mathbf{k}^{\prime}\mathbf{q}}V_{\mathbf{k}\mathbf{k}^{\prime}}
c^{\dag}_{\mathbf{k}\smash{+}\mathbf{q}/2\uparrow}c^{\dag}_{-\mathbf{k}\smash{+}\mathbf{q}/2\downarrow}
c_{-\mathbf{k}^{\prime}\smash{+}\mathbf{q}/2\downarrow}c_{\mathbf{k}^{\prime}\smash{+}\mathbf{q}/2\uparrow}
\end{eqnarray}
where $c_{\mathbf{k}\sigma}$ ($c^{\dag}_{\mathbf{k}\sigma}$) is the
annihilation (creation) operator for a fermion with momentum
$\mathbf{k}$, kinetic energy
$\xi_{\mathbf{k}\sigma}\smash{=}\epsilon_{\mathbf{k}}\smash{-}\mu_{\sigma}$,
and spin $\sigma$ ($=\uparrow,\downarrow$); $\mu_{\sigma}$ is the
chemical potential of each component, and
$\epsilon_{\mathbf{k}}\smash = \hbar^{2}k^{2}/2m$.

We consider condensate pairs with zero center-of-mass momentum, ($\mathbf{q}\smash{=}0$).
${\mathcal{H}}$ is mean-field (MF) decoupled via the
``spin''-triplet ($S=1,m_s=0$) pairing gap function
$\Delta_{\downarrow\uparrow}(\mathbf{k})\equiv
\Delta(\mathbf{k})\smash{=}\smash{-}\sum_{\mathbf{k}^{\prime}}
V_{\mathbf{k}\mathbf{k}^{\prime}}\langle
c_{-\mathbf{k}^{\prime}\uparrow}
c_{\mathbf{k}^{\prime}\downarrow}\rangle$
giving:
\begin{eqnarray}\label{Ham-MF}
{\mathcal{H}}^{\mathrm{MF}}&=&\sum_{\mathbf{k}\sigma}\xi_{\mathbf{k}\sigma}c^{\dag}_{\mathbf{k}\sigma}c_{\mathbf{k}\sigma}\nonumber\\
&-&\!\!\sum_{\mathbf{k}}
\left[\Delta(\mathbf{k})c^{\dag}_{\mathbf{k}\uparrow}
c^{\dag}_{-\mathbf{k}\downarrow}\smash{+}\mathrm{H.c.}\right]\nonumber\\
&-&\!\!\sum_{\mathbf{k}}\left|\Delta(\mathbf{k})\right|^2
/V_{\mathbf{k}\mathbf{k}}
\end{eqnarray}
To obtain the variational ground state energy of the MF Hamiltonian
(\ref{Ham-MF}) we use the equation-of-motion method \cite{LQ06} for
the imaginary-time ($\tau=it$) normal ($G_{\sigma\sigma^{\prime}}$)
and anomalous ($F_{\sigma\sigma^{\prime}}$) Green's functions:
\begin{eqnarray}
\partial_{\tau}G_{\sigma\sigma^{\prime}}(\mathbf{k},\tau)&=&\smash{-}\delta(\tau)\delta_{\sigma\sigma^{\prime}}
\smash{-}\xi_{\mathbf{k}\sigma}G_{\sigma\sigma^{\prime}}(\mathbf{k},\tau)\nonumber\\
&+&\Delta_{-\sigma\sigma}(\mathbf{k})
F_{-\sigma\sigma^{\prime}}(\mathbf{k},\tau),\label{G-normal}\\
\partial_{\tau}F_{\sigma\sigma^{\prime}}(\mathbf{k},\tau)&=&
\xi_{\smash{-}\mathbf{k}\sigma}
F_{\sigma\sigma^{\prime}}(\mathbf{k},\tau)\nonumber\\
&+&\Delta^{*}_{\sigma-\sigma}(\mathbf{k})
G_{-\sigma\sigma^{\prime}}(\mathbf{k},\tau).\label{G-anomal}
\end{eqnarray}
Eqs.\ (\ref{G-normal}) and (\ref{G-anomal}) are Fourier transformed
with $\tau\smash{\rightarrow}i\omega_{n}\smash{\equiv}\nu$ and
$\partial_{\tau}\smash{\rightarrow}\smash{-}\nu$, where
$i\omega_{n}\smash{=}(2n\smash{+}1)\pi T$ are Matsubara frequencies.
This gives $G_{\sigma\sigma}(\mathbf{k},\nu)={\smash{-}(\xi_{-\mathbf{k}-\sigma}\smash{+}\nu)}
/\left[{(\xi_{\mathbf{k}\sigma}\smash{-}\nu)
(\xi_{-\mathbf{k}-\sigma}\smash{+}\nu)
\smash{+}|\Delta_{\sigma,\smash{-}\sigma}(\mathbf{k})|^2}\right]$
and $F_{\sigma-\sigma}(\mathbf{k},\nu) \smash{=}
{\Delta^{*}_{\sigma-\sigma}(\mathbf{k})}/[(\xi_{\mathbf{k}-\sigma}-\nu) (\xi_{-\mathbf{k}\sigma}+\nu)
+|\Delta_{\sigma-\sigma}(\mathbf{k})|^2] \smash{=} - F_{-\sigma\sigma}(\mathbf{k},\nu)$.

We separate the radial and angular parts of the
interaction potential: $V_{\mathbf{k}\mathbf{k}^{\prime}}=
(4\pi/3)\sum_{m}V_{kk^{'}}Y_{1,m}(\hat{\mathbf{k}})
Y_{1,m}^{*}(\hat{\mathbf{k}^{\prime}})$ ($\hat{\mathbf{k}}$=$(\theta,\phi)$),
and take for $V_{kk^{'}}$ a separable form convenient for analytic calculations:
$V_{kk^{'}}=\lambda w(k)w(k^{\prime})$. Using $\langle
c_{-\mathbf{k}^{\prime}\downarrow}
c_{\mathbf{k}^{\prime}\uparrow}\rangle\equiv
F^{*}_{\downarrow\uparrow}(\mathbf{k}^{\prime},\tau\smash{=}0^{-})$
and Fourier transforming, we obtain for the gap function,
$\Delta(\mathbf{k})$=$-(1/k_BT)\sum_{\nu\mathbf{k}^{\prime}}
V_{\mathbf{k}\mathbf{k}^{\prime}}
F^{*}_{\downarrow\uparrow}(\mathbf{k}^{\prime},\nu)e^{\nu 0^{+}}$.
With the separable form
of $V_{kk^{\prime}}$, the gap function can be written as:
$\Delta(\mathbf{k})=\sum_{m}\Delta_{m}
w(k)Y_{1m}(\hat{\mathbf{k}})\equiv w(k)\Delta(\hat{k})$, where
$\Delta_{m}=-(1/k_BT)\sum_{\nu \mathbf{k}^{\prime}}\lambda
w(k^{\prime}) Y^{*}_{1m}(\hat{\mathbf{k}^{\prime}})
F^{*}_{\downarrow\uparrow}(k^{\prime},\nu)e^{\nu 0^{+}}$. With
$F_{\downarrow\uparrow}(k,\nu)$ known, we finally obtain the equations for the gap amplitudes ($m=0,\pm 1$),
\begin{equation}\label{gapeqn}
\Delta_{m}=-\!\lambda\!\sum_{\mathbf{k}}\! w(k)Y^{*}_{1m}
(\hat{\mathbf{k}})\Delta(\mathbf{k})[n_{F}(E^{-}_{k})\smash{-}n_{F}(E^{+}_{k})]/(2E_{k})
\end{equation}
where  $E^{\pm}_{k}=-h\pm E_{k}$,
$h=(\xi_{k\downarrow}-\xi_{k\uparrow})/2=(\mu_{\uparrow}-\mu_{\downarrow})/2$,
$E_{k}=[{\xi_{k}^{2}+|\Delta^{2}(\mathbf{k})}|]^{1/2}$, with
$\xi_{k}$=$(\xi_{k\uparrow}\smash{+}\xi_{k\downarrow})/2$=$\epsilon_{k}-\mu$
and $\mu$=$(\mu_{\uparrow}\smash{+}\mu_{\downarrow})/2$;
$n_F(E_k^{\pm})$ are Fermi functions. We can now average
(\ref{Ham-MF}) and calculate the variational ground-state energy
$E_{G}$. At $T$=$0$ and for $h$$>$$0$ this is given by:
\begin{eqnarray}
E_{G}&\equiv&\langle{\mathcal{H}}\rangle\!=\!\sum_{k}
\left\{\xi_{k}\smash{-}E_{k}
\smash{+}\frac{|\Delta(\mathbf{k})|^{2}}{2\epsilon_{k}}\right\}\nonumber\\
&+&\!\!\sum_{k}\left\{(E_{k}\smash{-}h)\theta(\smash{-}E_{k}\smash{+}h)\right\}
-\frac{1}{g}\sum_{m}|\Delta_{m}|^{2},\label{GroundE}
\end{eqnarray}
where the regularized interaction $g$ is given by
$1/g$=$1/\lambda$+$(1/(2\pi\hbar)^{3})\int w^2(k)
d^3\mathbf{k}/2\epsilon_{k}$.

{\it Ground State Analysis:} First, we carry out an analytic study of the p-wave superfluid ground state structure. We assume
$|\Delta(\mathbf{k})|\ll E_{k}$, and expand $E_k$ in powers of
$|\Delta(\mathbf{k})|$ keeping terms to 4th order: $E_{k}\sim
|\xi_k|
[1\smash{+}|\Delta(\mathbf{k})|^{2}/2\xi^{2}_{k}\smash{-}|\Delta(\mathbf{k})|^{4}/8\xi^{4}_{k}]$.
Substituting this expression for $E_{k}$ in (\ref{GroundE}) and
converting momentum sums to integrals we obtain a simple
representation for
$E_{G}=\alpha\int|\Delta(\hat{k})|^{2}d\Omega+\beta\int
|\Delta(\hat{k})|^{4}d\Omega+\gamma$, where
$\alpha=(1/(2\pi\hbar)^{3})\int_{E_{k}<h}
w^{2}(k)k^{2}dk/2\epsilon_{k}+(1/(2\pi\hbar)^{3})\int_{E_{k}>h}
w^{2}(k)k^{2}dk(1/2\epsilon_{k}\smash{-}1/2\xi_{k})-1/g$,
$\beta=(1/(2\pi\hbar)^{3})\int_{E_{k}>h}
w^{4}(k)k^{2}dk/8|\xi_{k}|^{3}$ and
$\gamma=(4\pi/(2\pi\hbar)^{3})\int_{E_{k}<h}(\xi_{k}\smash{-}h)k^{2}dk$.
Existence of a superfluid phase requires
$\alpha$$<$$0$, otherwise minimization of $E_G$ forces the gap
to vanishes. The polarization ($P=(n_\uparrow-n_\downarrow)/(n_\uparrow + n_\downarrow)$)
dependence of $E_G$ is contained in $\alpha, \beta, \gamma$, which
depend on $h, \mu, w(k)$, while $\alpha$ alone depends explicitly on
the coupling $g$. Owing to these dependences the gap parameters are
sensitive to changes in $P$.

For fixed $\mu$ and $h$, and $E_k$ given above, the ground state energy $E_{G}$ is
obtained by minimizing (\ref{GroundE}) with respect to the pairing
amplitudes $\Delta_m$.
Thus we arrive at an analytic expression
for the ground state energy \cite{CAL07}:
\begin{eqnarray}\label{simple}
E_{G}=-\frac{\alpha^{2}}{8\beta}+ \gamma +
2\beta\left(|\Delta_{0}|^{2}+|\Delta_{1}|^{2}
+|\Delta_{-1}|^{2}+\frac{\alpha}{4\beta}\right)^{2}\nonumber\\
+\beta(|\Delta_{0}|^{2}\smash{-}2|\Delta_{1}||\Delta_{-1}|)^{2}
\smash{+}4\beta(1\smash{-}t)|\Delta_{0}|^{2}|\Delta_{1}||\Delta_{-1}|
\end{eqnarray}
where $t$=$\cos{\theta}$, with  $\theta =
\cos^{-1}{(\Delta_{0}\Delta^{*}_{1},\Delta^{*}_{0}\Delta_{-1})}$ =
${(2\phi_{0}\smash{-}\phi_{1}\smash{-}\phi_{-1})}$; $\phi_m$'s
are the phases associated with the gap amplitudes $\Delta_m$. From
(\ref{simple}), we find that for a stable p-wave
superfluid phase to exist, the following conditions have to be
satisfied simultaneously
\begin{eqnarray}\label{conditions}
(a)&|\Delta_{0}|^{2}\smash{+}|\Delta_{1}|^{2}\smash{+}|\Delta_{-1}|^{2} &=\smash{-}\alpha/4\beta \nonumber\\
(b)&|\Delta_{0}|^{2}\smash{-}2|\Delta_{1}||\Delta_{-1}| &= 0\\
(c)&(1\smash{-}t) |\Delta_{0}|^{2}|\Delta_{1}||\Delta_{-1}| &= 0\nonumber
\end{eqnarray}
giving the GS {\it global minimum} energy
\begin{equation}\label{global}
E_{G}= -\alpha^{2}/8\beta + \gamma .
\end{equation}
\begin{figure}[t]
{\scalebox{0.85}{\includegraphics[clip,angle=0]{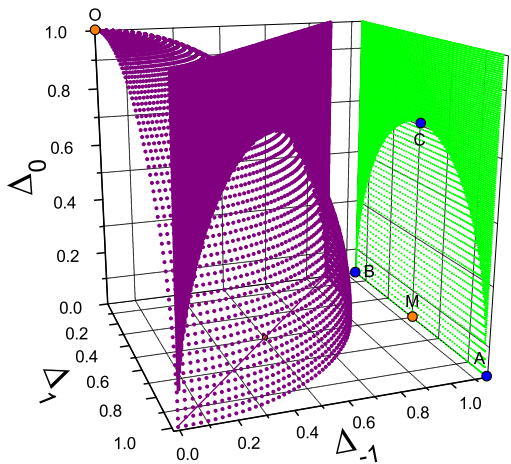}}}
\caption{(Color online) Geometric representation of p-wave
states in two-component population imbalanced system. States (e.g. A, B, C) exhibiting
global energy minimum lie on the semicircle formed by the
intersection of sphere surface with plane (see text). For clarity, plane containing
semicircle is also shown separately (green).
O (north pole) and M (on $\Delta_1$-$\Delta_{-1}$ plane) are states with local energy minimum.
$\Delta_m (m=0,\pm 1)$'s have been normalized to the sphere radius $\sqrt{- \alpha/4\beta}$; $\alpha < 0$.}
\label{Fig1}
\end{figure}
Conditions (a) and (b) together represent a semicircle formed by the
intersection of the surface of a sphere of radius $R$, $R^{2}=-\alpha/4\beta$,
with a plane defined by $|\Delta_{1}|+|\Delta_{-1}|$=$R$. All points on the semicircle
must necessarily satisfy the phase constraint (c). They represent a multitude of
``mixed'' superfluid states of the form $a\Delta_0 +b\Delta_1 +c\Delta_{-1}$.
A 3D geometric representation of the states is shown in Fig. 1.
The GS global energy minimum condition is satisfied only for $t=1$ in Eq. 8c, i.e. when the relative phase
$\theta = ((\phi_{0}\smash{-}\phi_{1})\smash{-}(\phi_{-1}\smash{-}\phi_{0}))= 2n\pi$.
For the states
$\Delta_{\pm 1}$, (points  A and B on the semicircle in Fig 1) this is always satisfied.
For the other states on this curve, containing varying admixtures of $\Delta_m$'s,
$E_G$ fluctuates with $\theta$, and attains the global minimum
only for $\theta=2n\pi$ (Fig. 3a); the fluctuation amplitude
depends on the specific state chosen; e.g. C at the top of the projected figure
has the maximum amplitude variation, but still lies lower than the
local minimum we discuss below.
$t$=$1$ corresponds to a parallel orientation of vectors
$\Delta_{0}\Delta^{*}_{1}$ and $\Delta^{*}_{0}\Delta_{-1}$; this
gives the same global minimum $E_{G}$=$-\alpha^{2}/8\beta+\gamma$
for three particular cases: (A)
$|\Delta_{0}|\smash{=}|\Delta_{-1}|$=$0$ and
$|\Delta_{1}|^2$=$-\alpha/4\beta$; (B)
$|\Delta_{0}|\smash{=}|\Delta_{1}|\smash{=}0$ and
$|\Delta_{-1}|^{2}$=$-\alpha/4\beta$; (C)
$|\Delta_{0}|^{2}$=$-\alpha/8\beta$ and
$|\Delta_{1}|^{2}\smash{=}|\Delta_{-1}|^{2}\smash{=}-\alpha/16\beta$.
These are shown as points A,B,C respectively in Fig. 1. In addition
to these, we also find the existence of {\it local} minimum with
$E_{G}^L$=$-\alpha^{2}/12\beta+\gamma$  for: (i)
$|\Delta_{0}|^{2}$=$-\alpha/4\beta$ and
$|\Delta_{1}|$=$|\Delta_{-1}|$=$0$; (ii) $|\Delta_{0}|$=$0$ and
$|\Delta_{1}|^{2}$=$|\Delta_{-1}|^{2}$=$-\alpha/12\beta$; these are
shown as O and M respectively in Fig. 1. Values for the GS energy
are completely determined by $\alpha$, $\beta$, and $\gamma$. In
evaluating these, we take the N-SR \cite{NOZ85} form:
$w(k)$=$k_{0}k/(k^{2}_{0}+k^{2})$, where $k_{0}$ is a cut-off
momentum.

In the limit of zero polarization, $P \rightarrow 0$, the GS energy
coefficient $\gamma \rightarrow 0$, while $\alpha<0, \beta>0$ remain
finite. Thus the GS structure of the polarized case is preserved for
zero polarization, with the same expressions for the energy global
and local minima (different numerical values). The states A and B
(equivalent to A by symmetry) above and in Fig. 1 correspond to the
finding of Ref.\ \cite{TIN05}. Our work reveals an additional set of
``mixed'' states on the intersecting semicircle (Fig. 1) that also
have the same global minimum for specific values of the relative
phase discussed above.

{\it Numerical Calculations:} Guided by our analysis of the p-wave
superfluid ground state, we carry out a detailed numerical study.
For fixed number densities, we solve self-consistently three gap
equations (Eq. 5) and two number equations given by:
\begin{equation}\label {numbereq}
n_{\sigma}=\sum_{k}\left<c_{k\sigma}^+c_{k\sigma}\right>
              =\sum_{k}G_{\sigma\sigma}(k,\tau=0^-),
\end{equation}
This gives us the gap amplitudes $\Delta_m$, and the chemical potentials $\mu_{\sigma}$.
Using these we obtain the ground state energy $E_G$ from Eq. 6, as well as
the coefficients $\alpha,\beta,\gamma$ that determine $E_G$ given by Eq. 7.
The agreement between our numerical $E_G$ and our analytical $E_G$ (based on an expansion of $E_k$)
is good to within 15\%.

We construct a polarization (P) - coupling ($1/k_F^3 a_t$) {\it phase diagram} in BEC-BCS
crossover regime (Fig. 2), enforcing the stability conditions:
that the stability matrix $(\partial^2{E_G}/\partial{\Delta_{m_i}}\partial{\Delta_{m_j}})$
is positive definite;  and that $\delta p/ \delta h >0$.
In Fig. 2, SF1 denotes a stable superfluid phase, corresponding to the states on
the `semicircle' in Fig. 1 that produce GS global minimum, i.e. with
relative phase $\theta$=$2n\pi$ among the gap parameters.
With increased polarization,
at $T$=0, SF1 becomes unstable, and gives way to the superfluid phase
SF2, corresponding to states with the local minimum discussed above.
In Fig. 2, states $\Delta_1$ and $\Delta_0$ were chosen in
SF1 and SF2 respectively; other choices of allowed global and local minimum states
give qualitatively the same phase diagram. At even larger polarizations,
SF2 becomes unstable to phase separation, PS. SF1, SF2, PS occupy a relatively much
narrower part of the phase diagram on the BCS side compared to the BEC side.
In our two-component system with inter-species interaction, PS persists into
full polarization, P=1.
This is reasonable because at P=1, the system is essentially a one-component
system in which the absence of minority species atoms makes inter-species interaction
inoperative. Such a system can exhibit superfluidity only under the effect of
intra-species interactions.
Our results suggest several interesting possibilities: A $T$=0 quantum phase transition
from SF1 to SF2 driven by polarization may be expected. Since the energies of SF2 and the
continuum of ``mixed'' states for $\theta \ne 2n\pi$ lie higher than the GS global
minimum, they may be accessed at finite-T.
Also, each p-wave state on the semicircle (Fig. 1) (generally
formed of a linear combination of the $\Delta_m$'s except
for the ``pure'' $\Delta_{\pm1}$ states A,B), is characterized by a
relative phase $2n\pi \le \theta \le (2n+2)\pi$. Hence, for each there is a multitude of
$\theta$-dependent states (Fig. 3a) with varying $E_G$ that may be interesting
to probe with experiments sensitive to such relative phases.
\begin{figure}[t]
{\scalebox{0.80}{\includegraphics[clip,angle=0]{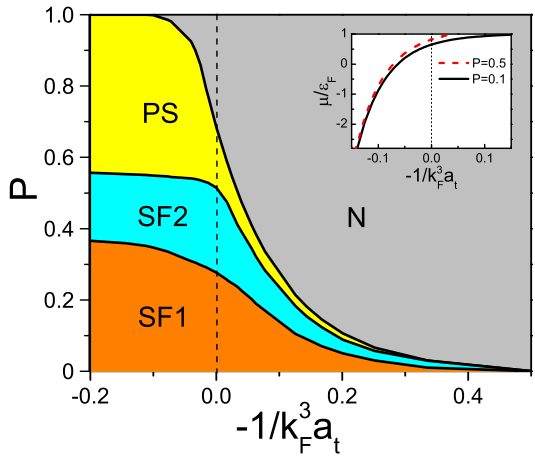}}}
\caption{(Color online) Polarization $P$ vs p-wave coupling $-1/(k^{3}_{F}a_{t})$
phase diagram of two-component one-channel Fermi system with $p$-wave pairing.
Shown are normal, N, and p-wave superfluid phases, SF1 and SF2, as well
phase separation, PS.
Unitarity limit is shown by the dashed vertical line. Inset:
Calculated chemical potential vs coupling across BCS and BEC regimes
for $P$ = 0.1 (solid line); 0.5 (dashed line).} \label{Fig2}
\end{figure}
The inset in  Fig. 2 shows the calculated behavior of
chemical potential $\mu$ across the  BEC-BCS regime.
It  deviates significantly from the Fermi energy
in a wider region around FR even on the BCS side, and drops much more rapidly to
negative values on the BEC side compared to the s-wave FR case.
For sufficiently weak coupling in the BCS regime, $\mu$ approaches
Fermi energy $\epsilon_{F}$.

Figs. 3b, 3c show the variation of $E_{G}$ with coupling $g$
(=$25\/k_F^3a_t/8\pi$) at fixed polarizations $P$, and with $P$ for
fixed $g$. $E_G$ is normalized to the Fermi energy $\epsilon_F$ =
$\hbar^2 k_F^2/2m$, with $2k_F^3=k_{F\uparrow}^3+k_{F\downarrow}^3$.
For a given $P$, $E_{G}$ becomes less negative as $g$ approaches
unitarity; the trend is more noticeable for smaller $P$'s.
For a given $g$, $E_G$ lie higher for smaller $P$, presumably
due to the lower majority-species band becoming progressively more
occupied with increasing $P$, thereby lowering $E_{G}$ with increasing polarization.
For small $P$, $E_G$ becomes less negative with increasing $g$. This trend is reversed
for $P \ge 0.3$.
\begin{figure}[t]
{\scalebox{0.80}{\includegraphics[clip,angle=0]{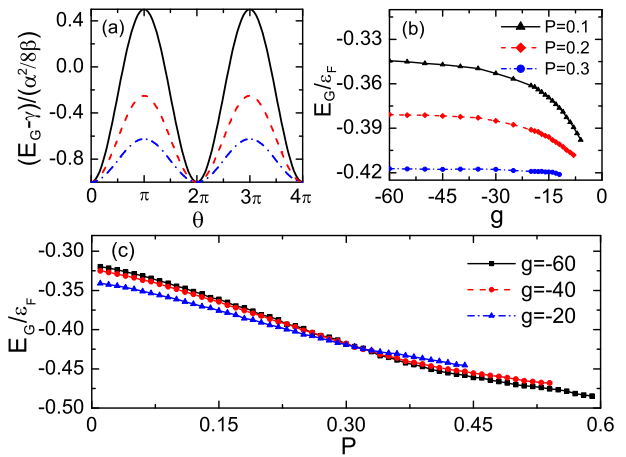}}}
\caption{(Color online) (a) Calculated ground state energy (scaled to global minimum) vs.
relative phase angle $\theta$ (see text).
Solid curve with maximal fluctuations of $E_{G}$ corresponds to
state C in Fig 1, ($|\Delta_{0}|^{2}$=$-\alpha/8\beta$,
$|\Delta_{1}|^{2}$=$|\Delta_{-1}|^{2}$=$-\alpha/16\beta$). The dash
and dot-dash curves correspond to two other states on the arc in
Fig. 1: $|\Delta_{0}|^{2}$=$-\alpha/11.3\beta$,
$|\Delta_{1}|^{2}$=$-\alpha/76.3\beta$,
$|\Delta_{-1}|^{2}$=$-\alpha/6.73\beta$, and
$|\Delta_{0}|^{2}$=$-\alpha/16\beta$,
$|\Delta_{1}|^{2}$=$-\alpha/186\beta$,
$|\Delta_{-1}|^{2}$=$-\alpha/5.49\beta$ respectively. (b) Calculated
ground state energy vs. coupling $g$ for different polarizations, P.
(c) Calculated ground state energy vs. polarization $P$ for
different coupling strengths $g$. } \label{Fig3}
\end{figure}

{\it Conclusions:} We believe we have presented new results and made several predictions
for two-component population imbalanced Fermi systems across the entire
BCS-BEC regimes. Future experiments in
cold fermions near p-wave FR may be able to provide tests of some of these
predictions. Insight into the nature of the orbital part of our superfluid states may be gained
from measuring the angular dependence of momentum distributions; from molecular spectroscopy
using light radiation; or possibly measurements of zero sound attenuation.
The work could also be of interest to other two-component Fermi systems where
p-wave intra-species couplings are small or negligible.

We acknowledge helpful discussions with Jason Ellis, Randy Hulet, Harry Kojima,
and Adriana Moreo. The work was partly supported by
funding from ICAM. One of us (F. Popescu) acknowledges a Fellowship from ICAM.

\suppressfloats
\end{document}